# Tevatron Collider Status and Prospects

Ronald S. Moore
*Fermi National Accelerator Laboratory, Batavia, IL 60510, USA*

The Tevatron proton-antiproton collider at Fermilab continues operation as the world's highest energy particle accelerator by delivering luminosity at a center-of-mass energy of 1.96 TeV. We review recent performance and plans for the remainder of Run 2.

## 1. INTRODUCTION

Fermilab's Tevatron began Run 2 operation as a proton-antiproton collider ($\sqrt{s}$ = 1.96 TeV) in March 2001. As of summer 2009, the Tevatron has delivered nearly 7 fb$^{-1}$ of luminosity to both of the CDF and D0 experiments – see Fig. 1. Although the Tevatron is currently scheduled to end operation October 2010, it may be extended to October 2011. Our goal is to deliver 9.3 fb$^{-1}$ (12 fb$^{-1}$) to the experiments by the 2010 (2011) end dates.

Since the previous long shutdown ended in November 2007 until June 2009, the Tevatron delivered over 3.5 fb$^{-1}$ while the record peak luminosity increased from 292 to 350 μb$^{-1}$/s (1 μb$^{-1}$/s = 10$^{30}$ cm$^{-2}$ s$^{-1}$) (Fig. 2). Over that same period, many weeks had integrated luminosities exceeding 50 pb$^{-1}$ and the record reached 73.1 pb$^{-1}$ delivered in a single week. (Fig. 1)

## 2. TEVATRON OPERATION

The Tevatron is a 1 km radius synchrotron with maximum beam energy of 980 GeV. The 36 proton and antiproton bunches are distributed into 3 trains of 12 bunches each with 396 ns bunch spacing. The beams circulate within a single beam pipe and electrostatic separators are used to kick the beams onto distinct helical orbits to keep the beams separated except at the collision points.

Increases in peak luminosity during Run 2 have arisen chiefly by higher beam intensities, especially antiprotons, and smaller beam size (lower $\beta^*$ at the interaction points and smaller injected beam emittances). Improvements in the Booster, slip-stacking in the Main Injector [1], and Antiproton Source [2] have led to increasing antiproton production rates. Electron cooling in the Recycler [3] has resulted in higher antiproton intensities with smaller emittances available for injection into the Tevatron.

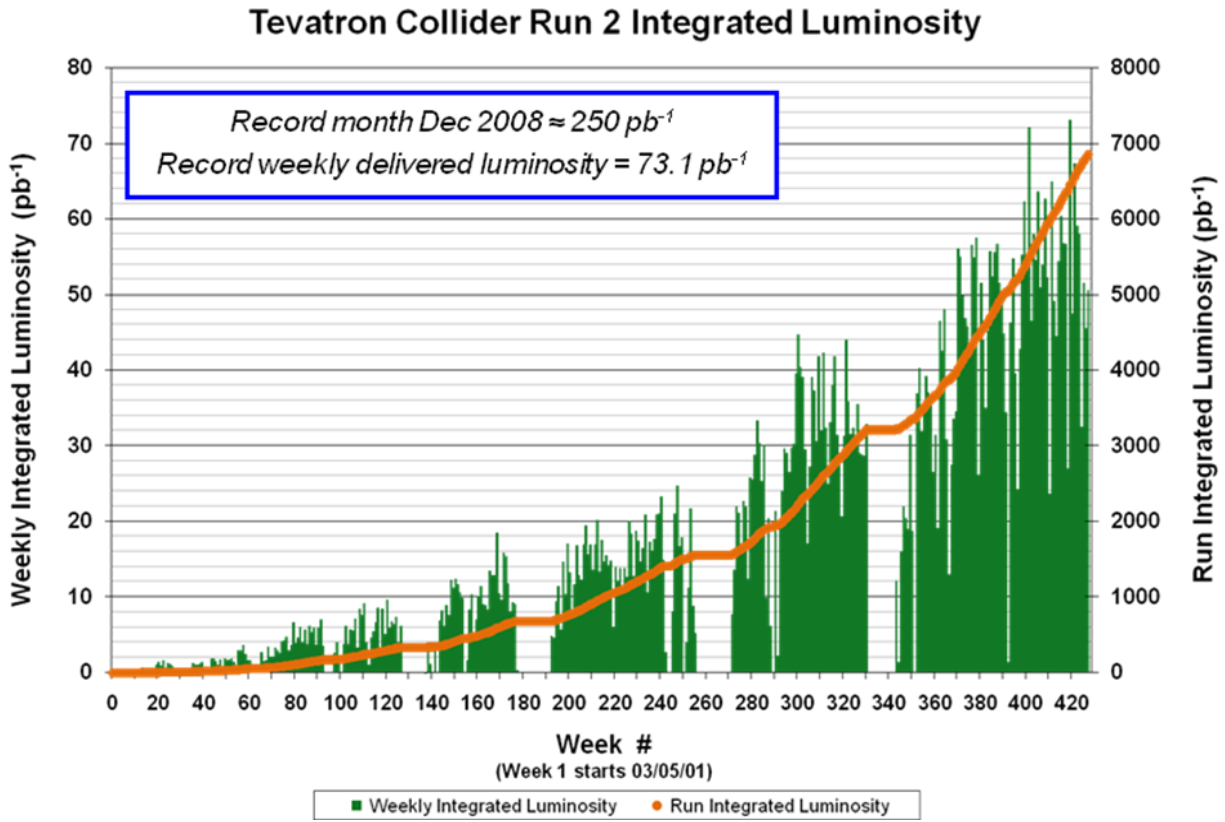

Figure 1: Weekly and total delivered luminosity delivered by the Tevatron over Run 2.

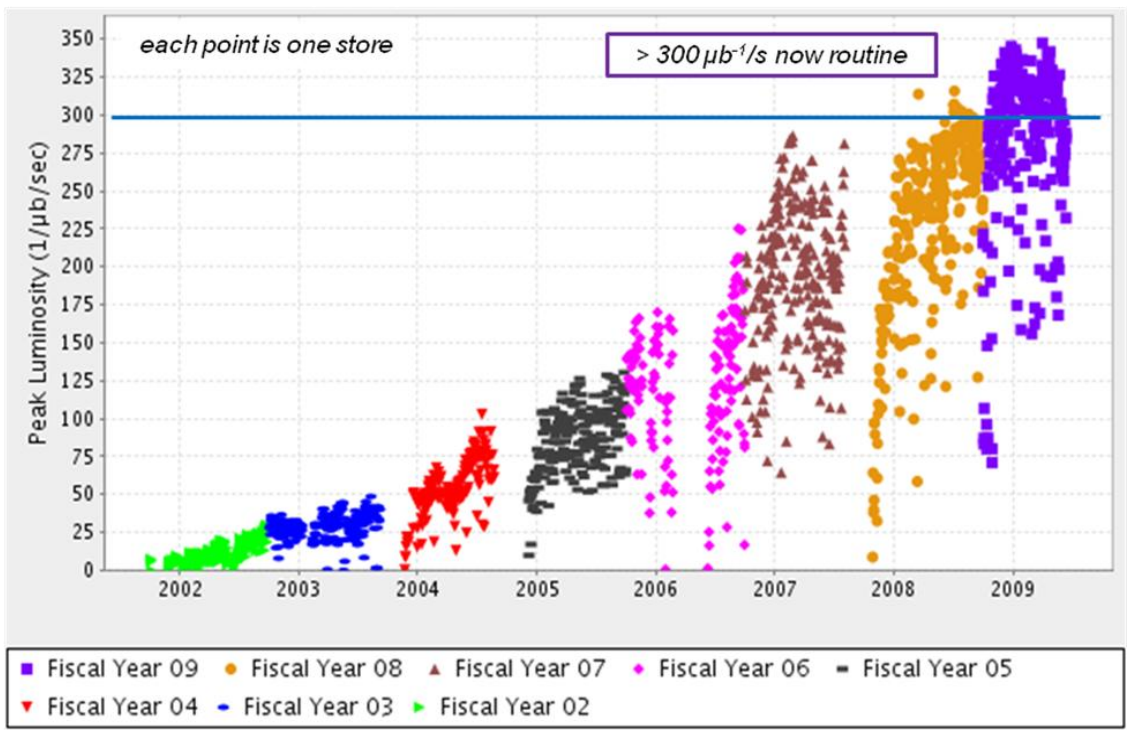

Figure 2: Initial luminosities for each Tevatron store from October 2002 – June 2009.

All of the major Tevatron Run 2 upgrades have been completed, so operational improvements are being pursued in order to maximize integrated luminosity delivered to the experiments. These efforts, highlighted in [4], include optimization of store length with antiproton accumulation rates, faster shot-setups, and additional automation of procedures (tuning and orbit smoothing) to allow greater reproducibility from store to store.

Since both proton and antiproton beams circulate within a single beam pipe, the electromagnetic field from one beam disrupts the other when they pass nearby or through each other. These detrimental beam-beam effects from both head-on collisions and long-range interactions at the parasitic crossing points play a major role in the overall machine performance [5]. We would like to maintain at least 5σ separation at the parasitic crossing points to reduce the long-range beam-beam effects, but it is not always possible. Beam-beam effects impact all phases of operation: injection, acceleration, low-beta squeeze and collisions.

The highest antiproton intensities can be problematic since the antiprotons can be a factor of 3-4 smaller transversely than the protons. High proton losses, exacerbated by the antiprotons, occasionally quench the Tevatron during the squeeze. To combat such losses, we have started deliberately "blowing-up" the antiproton emittances at 980 GeV prior to the low-beta squeeze. An electronic noise source on a directional stripline [6] enlarges the antiproton emittances ≈25% in order to reduce the beam-beam effects.

The (linear) head-on beam-beam tune shift parameter for each beam during collisions is given by

$$\xi = \frac{3r_0}{2}\frac{N}{\varepsilon}$$

where $r_0$ is the classical proton radius, N and ε are the number of particles and transverse emittance of the incoming colliding bunch, respectively. Plugging in typical values of N and ε for protons ($250 \times 10^9$/bunch, 18 pi mm mrad) and antiprotons ($70 \times 10^9$/bunch, 5 pi mm mrad) gives ≈0.020 for both beams (total from both interaction points) – remarkable for a proton-antiproton collider that both beams have similarly large tune shift.

The transverse emittance mismatch between the protons and antiprotons also causes a significant difference in the beam-beam induced tune spread. Fig. 3 illustrates that the antiproton tune shift caused by the larger protons is peaked near the value of the head-on beam-beam tune shift parameter; however, the protons have a rather broad, nearly uniform, tune shift distribution.

The proton tune spread makes it very difficult to avoid tune resonances for all protons in a bunch. Consequently, the protons tend to suffer more non-luminous beam loss than the antiprotons. (Non-luminous loss simply means loss from processes other than burn-up in useful collisions.) This large difference in tune spread is another motivation to obtain more equal beam emittances either by enlarging the antiproton emittances like we already do or injecting smaller emittance protons.

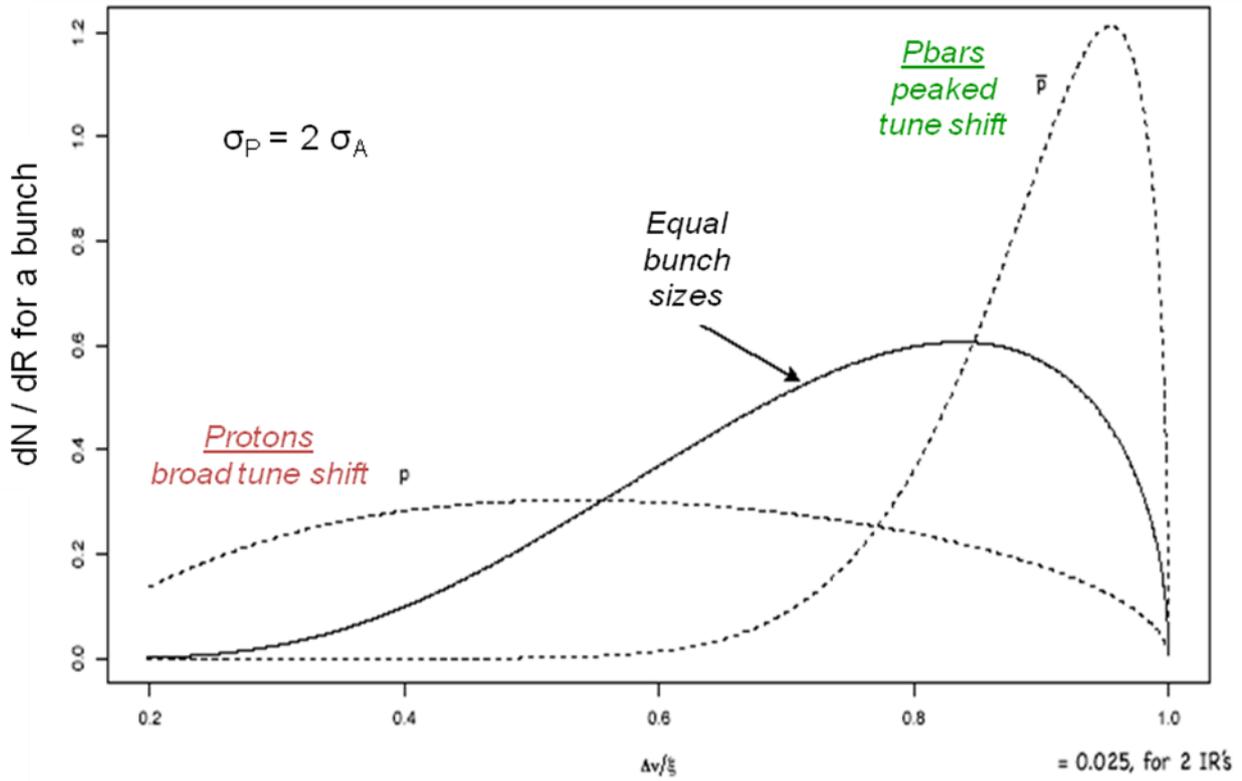

Figure 3: Distributions of beam-beam tune shift (expressed relative to the head-on beam-beam tune shift parameter) for protons and antiprotons where protons have transverse emittance twice that of the antiprotons (dashed lines) or both beams have the same emittance (solid line).

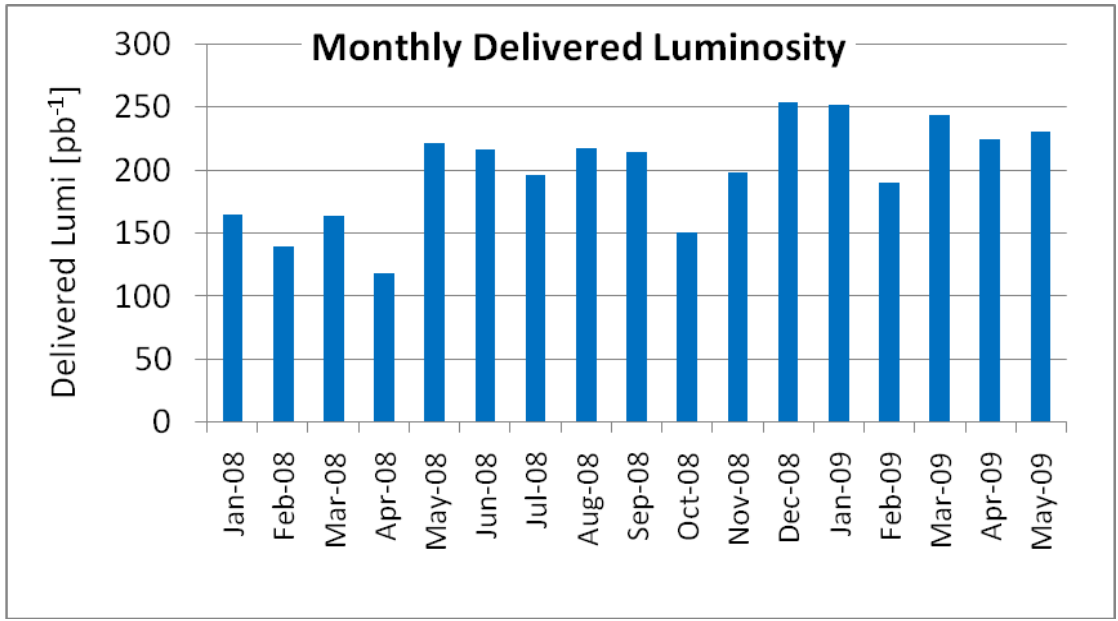

Figure 4: Tevatron delivered luminosity by month from January 2008 – May 2009.

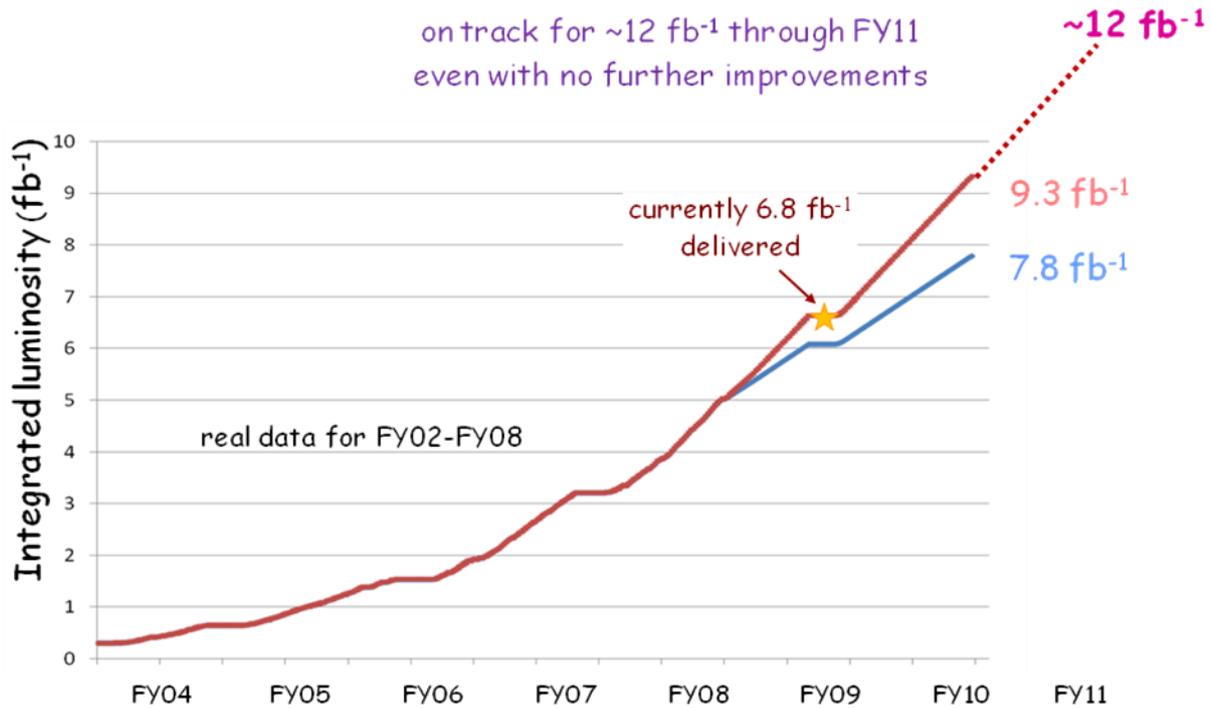

Figure 5: Projection for Tevatron Run 2 delivered luminosities through FY(fiscal year) 2010 and FY 2011.

## 3. PROJECTIONS

Fig. 4 demonstrates that the Fermilab accelerator complex can support routine operation of the Tevatron delivering 200 pb$^{-1}$/month. There will be a three month shutdown during summer 2009 for maintenance, installation of new Booster correctors, and Main Injector construction for the NOvA project. Following the shutdown and several weeks of restart commissioning, our plan is to resume delivering luminosity at that rate (hopefully surpassing it) with a goal of 2.5 fb$^{-1}$/year.

Fig. 5 shows the projection of integrated luminosity over the remainder Run 2. Given the 6.8 fb$^{-1}$ already delivered prior the 2009 shutdown, we expect to have delivered over 9 fb$^{-1}$ by the end of FY 2010 and nearly 12 fb$^{-1}$ if Run 2 extends to the end of FY 2011.

### Acknowledgments

The author thanks the Fermilab staff of the Accelerator Division and Accelerator Physics Center for their dedicated support to make Tevatron Collider Run 2 successful.

This work is supported by Fermi Research Alliance, LLC under Contract No. DE-AC02-07CH11359 with the United States Department of Energy.